ARTICLE  OPEN

# Lattice frustration in spin-orbit Mott insulator $Sr_3Ir_2O_7$ at high pressure

Jianbo Zhang[1], Dayu Yan[2], Sorb Yesudhas[1], Hongshan Deng[1], Hong Xiao[1], Bijuan Chen[1], Raimundas Sereika[1], Xia Yin[1], Changjiang Yi[2], Youguo Shi[2], Zhenxian Liu[3], Ekaterina M. Pärschke[4], Cheng-Chien Chen[4], Jun Chang[5], Yang Ding[1] and Ho-kwang Mao[1,6]

The intertwined charge, spin, orbital, and lattice degrees of freedom could endow 5$d$ compounds with exotic properties. Current interest is focused on electromagnetic interactions in these materials, whereas the important role of lattice geometry remains to be fully recognized. For this sake, we investigate pressure-induced phase transitions in the spin-orbit Mott insulator $Sr_3Ir_2O_7$ with Raman, electrical resistance, and x-ray diffraction measurements. We reveal an interesting magnetic transition coinciding with a structural transition at 14.4 GPa, but without a concurrent insulator-metal transition. The conventional correlation between magnetic and Mott insulating states is thereby absent. The observed softening of the one-magnon mode can be explained by a reduced tetragonal distortion, while the actual magnetic transition is associated with tilting of the $IrO_6$ octahedra. This work highlights the critical role of lattice frustration in determining the high-pressure phases of $Sr_3Ir_2O_7$. The ability to control electromagnetic properties via manipulating the crystal structure with pressure promises a new way to explore new quantum states in spin-orbit Mott insulators.

## INTRODUCTION

Exotic ground states in 5$d$ quantum materials, such as spin liquids,[1,2] Weyl semimetals,[3] topological insulators,[4,5] and superconductors,[6,7] are commonly believed to be driven by the competition between electron interaction $U$, spin-orbit coupling (SOC) $\xi$, and hopping amplitude $t$. In particular, for $Ir^{5+}$ atoms in octahedral crystal field, the 5$d$ orbital degeneracy is lifted by the strong on-site SOC, and the ground state is formed by effective $S = 1/2$ pseudospins.[8] The lattice degree of freedom and its coupling to electron orbital angular momenta therefore have long been thought to play a trivial role.[9]

However, recently it has been recognized that pseudospin-lattice coupling may have strong impact on the low-energy physics of 5$d$ materials.[10] In the single-layer perovskite $Sr_2IrO_4$, canted $S = 1/2$ pseudospins in the antiferromagnetic phase are shown to rigidly lock to $IrO_6$ octahedra due to strong SOC, and the pseudospin orientations rotate together with the octahedra under applied electric current.[11] Moreover, pseudospin-lattice coupling can induce a tetragonal-to-orthorhombic structural transition and explain the in-plane magnon gaps of $Sr_2IrO_4$.[12] Jahn–Teller effect also can explain some high energy features of different iridates in resonant inelastic x-ray scattering (RIXS),[13] and the avoidance of metallization of $Sr_2IrO_4$ under pressure.[14] These findings suggest that subtle structural changes may influence critically the low-energy Hamiltonian. While determining the exact role of lattice variable in 5$d$ materials remains a challenge, applying pressure opens up an avenue for such research, since it could possibly decouple entangled degrees of freedom during phase transitions.[15]

In this work, we apply pressure to the double-layered perovskite $Sr_3Ir_2O_7$, which is the middle member of the Ruddlesden-Popper series $Sr_{n+1}Ir_nO_{3n+1}$ ($n = 1, 2, \infty$). This material provides an interesting playground to study phase transitions, as it is considered in the weak Mott limit[16] with a relatively small charge gap and tiny magnetic moment.[17] $Sr_3Ir_2O_7$ has been extensively studied recently,[8,16–31] and its crystal structure is commonly reported to stabilize in $I4/mmm$ symmetry.[18] However, when the rotation angle $\alpha$ (describing in-plane $IrO_6$ rotation, ~11–12°) and the tilt angle $\beta$ (describing the deviation of $IrO_6$ $c$-axis from the $z$-direction, ~less than 1°) are taken into account, the symmetry is reduced to $Bbca$,[17] or $Bbcb$,[19] or even $C2/c$.[20] Furthermore, the $IrO_6$ octahedron itself has a slight tetragonal distortion. Therefore, $Sr_3Ir_2O_7$ serves as an ideal candidate for studying how lattice frustrations affect the electromagnetic properties and structural stability at high pressure.

## RESULTS AND DISCUSSION

Figure 1 shows the temperature-dependent Raman spectra of $Sr_3Ir_2O_7$ at ambient pressure. Six Raman modes are identified, respectively, at 146, 181, 269, 392, 592, and 1360 cm$^{-1}$ based on the room-temperature spectrum. The frequencies are close to those reported in the literature.[21] As temperature decreases, a new Raman peak appears at $780 \pm 8$ cm$^{-1}$ (96.7 ± 1.4 meV) at 270 K. We assign this new peak to the one-magnon mode, as its peak intensity decreases significantly with increasing temperature, and its energy is also consistent with the zone-center magnon excitation (~90 meV) reported in RIXS studies of $Sr_3Ir_2O_7$.[22,23] The one-magnon peak exhibits an asymmetric Fano-like line shape, which could result from interaction with a continuum of excitations. Above $T_N$ ~285 K,[17] the magnon peak disappears.

[1]Center for High-Pressure Science and Technology Advanced Research, Beijing 100094, People's Republic of China; [2]Institute of Physics, Chinese Academy of Sciences, Beijing 100190, People's Republic of China; [3]Institute of Materials Science, Department of Civil and Environmental Engineering, The George Washington University, Washington, DC 20052, USA; [4]Department of Physics, University of Alabama at Birmingham, Birmingham, AL 35294, USA; [5]College of Physics and Information Technology, Shaanxi Normal University, Xi'an 710119, People's Republic of China and [6]Geophysical Laboratory, Carnegie Institution of Washington, Argonne, IL 60439, USA
Correspondence: Jianbo Zhang (jianbo.zhang@hpstar.ac.cn) or Jun Chang (junchang@snnu.edu.cn) or Yang Ding (yang.ding@hpstar.ac.cn)

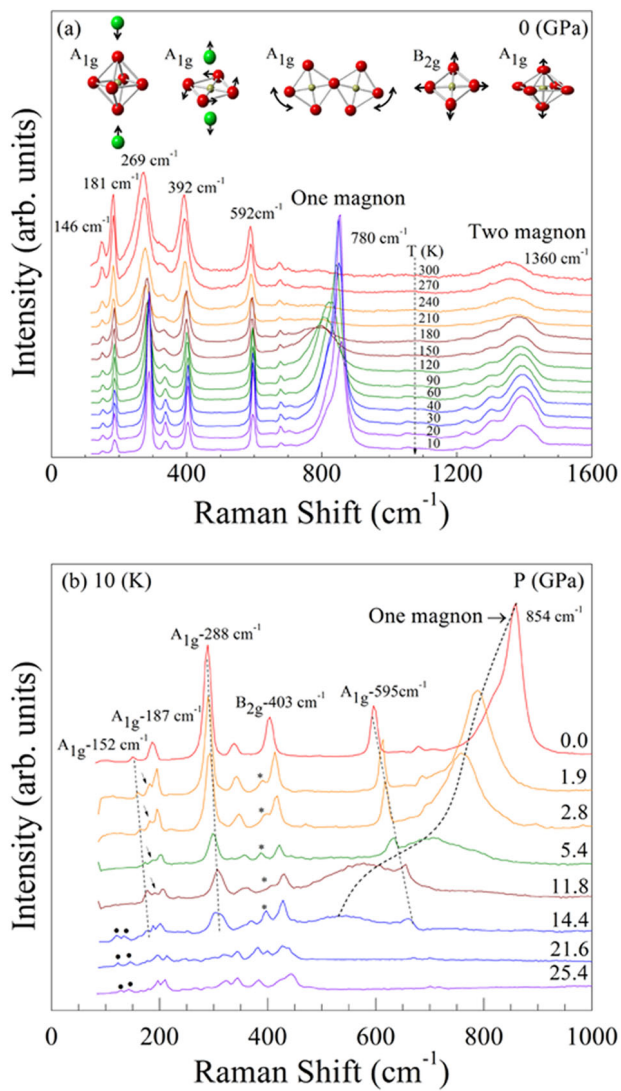

**Fig. 1** **a** Temperature-dependent Raman spectra of $Sr_3Ir_2O_7$ at ambient pressure. The direction of decreasing temperature is indicated by the dashed arrow line. Five major vibrational modes are identified, and their atomic motions in the polyhedral are represented at the top of the figure. **b** Raman spectra of $Sr_3Ir_2O_7$ for selected pressures at 10 K. The dotted lines are guides to the eye, showing the evolution of three $A_{1g}$ modes, respectively, at 152, 288 and 595 $cm^{-1}$. The dashed curve tracks the pressure dependence of the one-magnon peak. Since the two-magnon mode merges with the first order Raman mode of diamond, we cannot study its high-pressure behavior. Two additional phonon modes appearing at 14.4 GPa are indicated by dots. Additional modes occurring at 1.9 GPa close to the $A_{1g}$ (187 $cm^{-1}$) and $B_{2g}$ (403 $cm^{-1}$) modes are indicated by arrow and asterisk, respectively

The Raman spectra also reveal a two-magnon mode at 1360 ± 2.3 $cm^{-1}$ (168.62 ± 0.3 meV) [Fig. 1a], which is different from the previously reported energy ~1500 $cm^{-1}$ (~185 meV) measured by 632.8 nm laser.[21] The two-magnon peak increases slowly with decreasing temperature, and its intensity remains appreciable even at room temperature, suggesting a short-range spin-correlation character. In contrast, the previously reported two-magnon mode was strongly suppressed with temperature and vanished at $T_N$ = 285 K.[21] Given that the two-magnon mode in our study also could be excited by 488 nm laser, different excitation source is probably not the reason for the observed discrepancy. Rather, the discrepancy could originate from subtle variation in the compositions of different samples.

Apart from the assigned magnon modes at 780 and 1360 $cm^{-1}$, the other five Raman peaks are assigned to phonon modes, which could be indexed according to the $I4/mmm$ space group of tetragonal symmetry.[18] The fourteen Raman active modes expected from group theory include $\Gamma_{Raman} = 5A_{1g} + 2B_{1g} + 1B_{2g} + 6E_g$.[24] However, only one $B_{2g}$ mode at 392 $cm^{-1}$ and four $A_{1g}$ phonon modes are observable in our Raman experiments, whereas the rest of the modes are absent probably due to their weak Raman scattering cross sections.

As previously reported, the $A_{1g}$ mode at 146 $cm^{-1}$ corresponds to the stretching of Sr atoms against the $IrO_6$ octahedra.[24] The $A_{1g}$ mode at 181 $cm^{-1}$ involves displacements of Sr atoms along the c-axis with antiphase motion of adjacent layers and in-plane rotations of O atoms.[21] The $A_{1g}$ mode at 269 $cm^{-1}$ is attributed to the bending of the Ir–O–Ir bond[21,24] due to $IrO_6$ rotation, and the $B_{2g}$ mode at 392 $cm^{-1}$ is associated with some out-of-plane atomic displacement of in-plane oxygens.[21] Finally, the $A_{1g}$ mode at 592 $cm^{-1}$ originates from the vibration of apical oxygen atoms in $IrO_6$ octahedra.[32] These modes are illustrated in Fig. 1a. When cooled down to 10 K, these Raman modes shift to higher frequencies as shown in Fig. 1b.

Figure 1b shows the pressure-dependent Raman spectra of $Sr_3Ir_2O_7$ at 10 K. First, we note that two weak phonon modes emerge close to the $A_{1g}$ mode (187 $cm^{-1}$) and $B_{2g}$ mode (403 $cm^{-1}$) at 1.9 GPa. These two emergent modes, indicated by arrow and asterisk, respectively, in Fig. 1b, have been discovered previously[24] and are visible below $T_N$ owing to magnetic interaction. Second, the one-magnon peak softens continuously and broadens with increasing pressure; the peak eventually disappears around 14.4 GPa, evincing a magnetic transition. In addition, two new modes appear below 152 $cm^{-1}$ and the above $A_{1g}$ mode weakens at around 14.4 GPa. Similarly, the $A_{1g}$ mode at 288 $cm^{-1}$ (the bending of the Ir–O–Ir bond) and 595 $cm^{-1}$ (the vibration of apical oxygens in $IrO_6$ octahedra) harden with pressure prior to almost vanishing around 14.4 GPa, which indicates increased $IrO_6$ rotation and tilt angles at high pressure. These changes in phonon modes suggest a structural transition concurring with the magnetic transition around 14.4 GPa. We also performed room temperature and high-pressure Raman measurements to provide further evidence of the structural transition at 23.2 GPa. The results are given in the Supplementary Material in Fig. S2.

The pressure-induced structural phase transition is also confirmed by X-ray diffraction (XRD) at room temperature. The XRD measurements are performed on single crystals, and the results confirm that the sample is stable in an $I4/mmm$ phase up to 21.2 GPa at room temperature, while the new phase was fitted with space group $C2$ (Fig. 2b). The integrated XRD patterns of $Sr_3Ir_2O_7$ up to 33.2 GPa are presented in the Supplementary Material in Fig. S3a. At 33.2 GPa, the X-ray diffraction pattern fitting using the symmetry of $I4/mmm$ starts to fail (see Fig. S3b in the Supplementary Material), suggesting the structures of the new phase should adopt a lower symmetry, i.e., $C2$. Detailed XRD analysis of the $Sr_3Ir_2O_7$ crystal performed to exclude possible admixture of $Sr_2IrO_4$ is also shown in the Supplementary Material in Fig. S4.

Our discovery of magnetic and structural transitions at 14.4 GPa now could explain the mysterious origin of a second-order phase transition reported by Zhao et al.[25] In their work, the second-order transition derived from the P-V data was attributed to an insulator-metal like transition at ~13 GPa,[26] but they also suspected that magnetic transition might be a possible cause.[25] To investigate if there is a concurring insulator-metal transition (similar to that observed by in pump-probe experiment at ambient conditions[27]), we also perform electrical transport measurement on a $Sr_3Ir_2O_7$ single crystal. The results are plotted in Fig. 3a.

The electrical resistance within the *a-b* plane follows an activation law $R_{a-b}(T) = exp\,(\Delta/2k_BT)$, where $\Delta$ is the charge gap and $k_B$ is the Boltzmann's constant.[26,33] We obtain the value of $\Delta$ at each pressure point from linear fitting of $lnR\,(T)$ vs. $1/T$. The gap energy (black square) as a function of pressure is plotted in Fig. 3b.

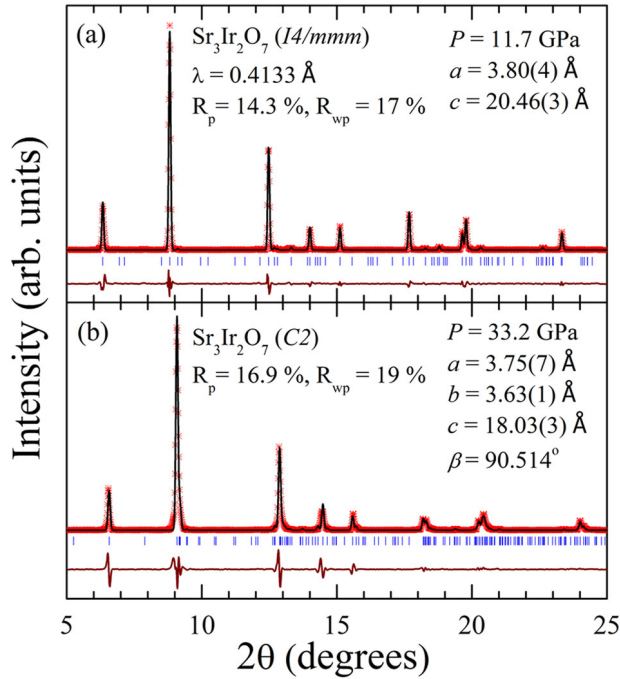

**Fig. 2** **a** Le Bail fitting of the tetragonal *I4/mmm* phase of $Sr_3Ir_2O_7$ at 11.7 GPa. **b** Le bail fitting of the higher-pressure monoclinic *C2* phase of $Sr_3Ir_2O_7$ at 33.2 GPa. The lattice parameters of each phase are written, respectively, at the top-right corner of each panel. In **a** and **b** the red circles are background subtracted data points, the black solid line presents the Le Bail fitting, blue tick marks are the calculated Bragg reflections, and the residuals of the fitting are shown by the black line beneath

From this analysis, $Sr_3Ir_2O_7$ remains an insulator even when the structural and magnetic transitions concur at 14.4 GPa. It is expected to metalize at 55.6 GPa, which is close to the critical pressure reported previously.[26,28,29] Pressure thus decouples the insulator-metal transition from the magnetic and structural transitions in $Sr_3Ir_2O_7$, while these transitions remain coupled in pump-probe measurements.[27]

To further address how lattice frustration affects the magnetic order, we use spin-wave theory to study the magnon excitations. In particular, we adopt the magnetic exchange Hamiltonian from the reference,[22] which describes well the spin-wave dispersion of $Sr_3Ir_2O_7$ in RIXS measurements. The pressure evolution of magnon dispersion could originate from a change in three microscopic parameters: the ratio of Hund's coupling to the on-site Coulomb interaction $\eta$ ($=J_H/U$), the $IrO_6$ rotation angle $\alpha$, and the effective tetragonal distortion $\theta$ that parametrizes the tetragonal splitting of $t_{2g}$ levels. It is important to understand which parameter dominates the softening of single-magnon energy under pressure.

First, pressure could enhance $\eta$ via screening the Hubbard $U$ while leaving $J_H$ nearly unchanged. Since we do not observe any metallization at 14.4 Gpa, the impact of pressure on $\eta$ should be small. Second, it has been shown that pressure can increase the rotation angle $\alpha$ by a few degrees.[28] However, our numerical calculations indicate that such a small change in $\alpha$ only lowers the magnon frequency by a few percent (see Fig. 4), which is not enough to account for the experimentally observed magnon softening. On the other hand, decreasing $\theta$ can significantly reduce the zone-center magnon energy and lift the degeneracy of the magnon branches (see Fig. S1 in the Supplementary Material). The evidence for reduced tetragonal distortion at high pressure is indeed found in the RIXS experiment by Ding et al.,[28] in which the peak width of spin-orbiton excitation at ~0.5 eV reduces from 0.56 eV at 0.98 GPa to 0.48 eV at 12.4 GPa. Such a reduction strongly suggests a reduced tetragonal distortion, as in principle the peak width should have increased under pressure due to the broadening of $J_{eff} = 1/2$ and $J_{eff} = 3/2$ bands. We thus conclude that one major effect of pressure on the material is to reduce the anisotropy arising from tetragonal distortion of $IrO_6$ octahedra (see Fig. 4).

Although the model could explain the magnon softening, it is unable to illustrate the disappearance of magnon. To explain this effect, the $IrO_6$ tilting angle $\beta$ should be considered. It already has been reported that the actual symmetry of $Sr_3Ir_2O_7$ should be $C2/c$ with a tilting angle $\beta$ less than 1 degree.[20] Such a small tilting usually has been regarded as having a trivial effect. In our high-pressure study, however, the tilting appears to be important. In fact, the observed blue shift and disappearance of the 152 cm$^{-1}$ and 595 cm$^{-1}$ Raman modes suggest that pressure enhances both α and β, which could induce buckling or disordering of IrO$_6$ octahedra along the c-axis and results in a local symmetry breaking (structural transition). The antiferromagnetic order (that gives rise to the magnon dispersion) in Sr$_3$Ir$_2$O$_7$ stems from the interlayer exchange coupling $J_c$ between Ir atoms mediated through apical oxygens.[30] A strong IrO$_6$ tilting could suppress $J_c$ and result in the disappearance of magnon excitation. In principle, full suppression of the magnon peak also could arise from other effects, such as staggered distortions where the intralayer interactions change differently in two neighboring layers. Spin-phonon coupling mediated by single-ion anisotropy also can induce a small in the spin gap, although the phonon renormalization after magnetic transition in Sr$_3$Ir$_2$O$_7$ appears weaker compared with that in other 5d compounds, such as Cd$_2$Os$_2$O$_7$.[34–38] A future comprehensive theoretical investigation of these effects is greatly needed.

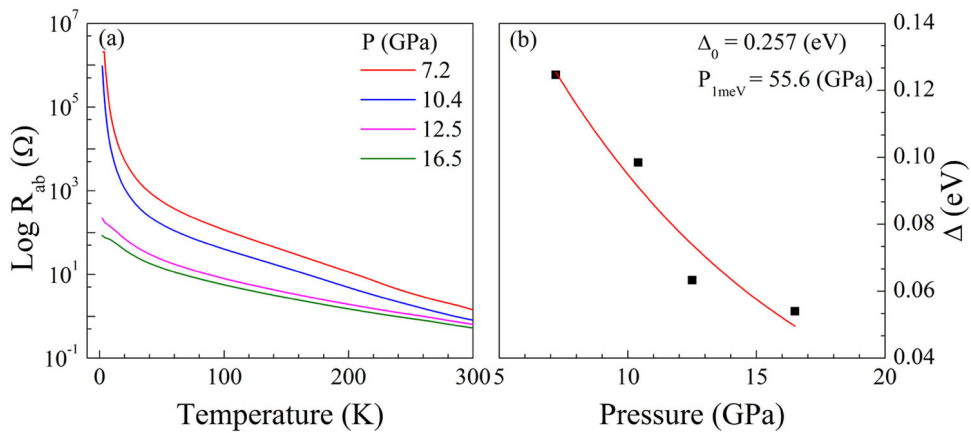

**Fig. 3** **a** Temperature-dependent electrical resistance of $Sr_3Ir_2O_7$ along the *a-b* plane ($R_{ab}$) plotted in log scale for selected pressures. $Sr_3Ir_2O_7$ remains in an insulating state even after the magnetic transition at ~14.4 GPa. **b** Pressure (*P*) vs. charge gap energy ($\Delta$) of $Sr_3Ir_2O_7$. The critical pressure for reaching the metallic state can be estimated by the following empirical relation: $\Delta(P) = \Delta_0\,exp(-\chi P)$, where $\Delta_0$ is the gap energy at ambient-pressure and $\chi$ determines the rate of decrease of $\Delta$ with *P*. A fitting (red line) using the above equation results in $\chi = 0.09976\,GPa^{-1}$ and $P_{1meV} = 55.6\,GPa$. This critical pressure 55.6 GPa for metallization is close to that reported in previous experiments.[26,28,29]

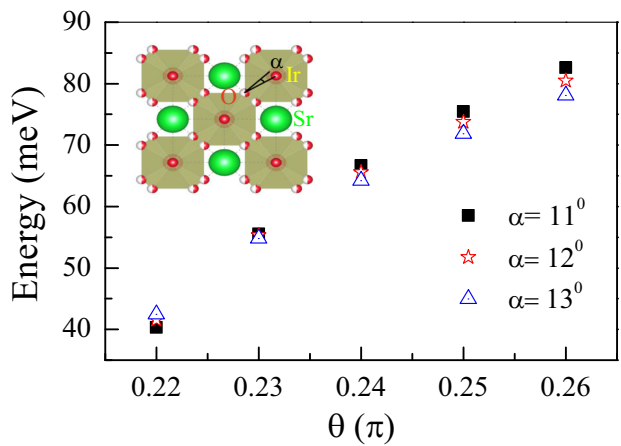

**Fig. 4** Spin-wave theory prediction for the one-magnon energy at the Brillouin zone center as a function of tetrahedral distortion (θ) for selected IrO$_6$ octahedron rotation angles (α). The ratio of Hund's coupling to the on-site Coulomb interaction $\eta = (J_H/U)$ is fixed. The theoretical model and parameters for ambient pressure are adopted from the ref. [22]

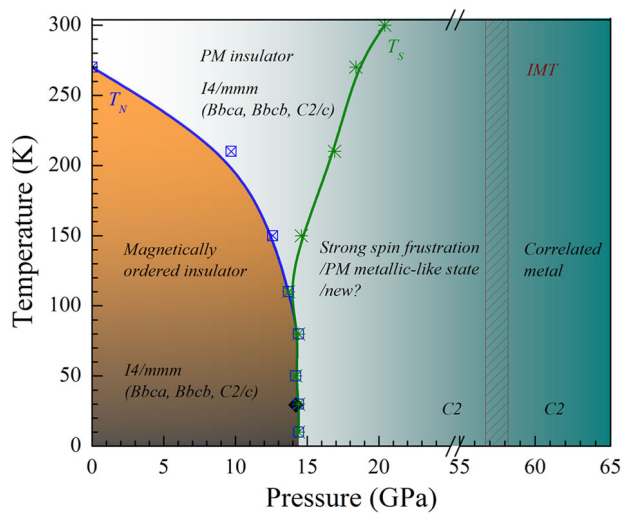

**Fig. 5** Schematic P-T phase diagram of Sr$_3$Ir$_2$O$_7$. The phase boundary ($T_N$) of paramagnetic (PM) insulator phase and magnetically ordered insulator phase and also the structural phase transition boundary ($T_S$) are based on our Raman data (squares and asterisks on the phase boundary). The black dot on the $T_N$ (or $T_S$) phase boundary represents the structural phase transition reported in the reference.[25] The insulator to metal transition (IMT) region is taken from the references.[26,28,29]

Finally, we summarize our findings in a phase diagram shown in Fig. 5. After the magnetic transition at 14.4 GPa and 10 K, Sr$_3$Ir$_2$O$_7$ is still an insulator. It remains challenging to characterize the magnetic structure and the underlying mechanism of phase transition. Based on similar phenomena observed in doping experiments, the insulating phase above 14.4 GPa could be paramagnetic, spin frustrated,[31] or even a totally new quantum state that has never been reported. Sr$_3$Ir$_2$O$_7$ at ambient conditions is located approximately at the center of the phase diagram plotted as functions of $U/t$ and $\xi/t$.[9] When pressure increases, both $U/t$ and $\xi/t$ could decrease, and the system could eventually reach the metallic region. We have reported previously an insulator-metal transition in Sr$_3$Ir$_2$O$_7$ around 55–59.5 GPa,[28] which is consistent with other studies.[26,29] According to Fig. 5, the pathway for such a pressure evolution could possibly pass though the axion insulator phase[3] before it reaches the metallic regime. Therefore, it is not impossible that the insulating phase discovered here might be an axion state. We speculate that applying pressure could be a promising way to search for various predicted phases in 5d spin-orbit materials.[9]

In summary, we have revealed an interesting magnetic transition that concurs with a structural transition around 14.4 GPa by Raman measurements on Sr$_3$Ir$_2$O$_7$. We attribute the origin of these transitions to pressure-enhanced rotation and tilt angles of IrO$_6$ octahedra. The magnetic transition is shown to decouple from insulator-metal transition. The absence of usual correlation between magnetic and insulating phases in Sr$_3$Ir$_2$O$_7$ is similar to that in Sr$_2$IrO$_4$.[39,40] Our high-pressure study together with previous discoveries manifests the critical role of lattice frustration in determining the ground state properties of Sr$_3$Ir$_2$O$_7$, and maybe generically of 5d materials. Our work calls for more theoretical studies to unravel the interplay of intertwined degrees of freedom in spin-orbit systems and the exact mechanism of their phase transitions.

## METHODS

### Sample synthesis
Sr$_3$Ir$_2$O$_7$ single crystal was grown from flux method. High-purity SrCO$_3$, IrO$_2$ and SrCl$_2$·6H$_2$O powders were mixed together and placed in platinum crucible. The molar ratios of the source materials was 2:1:20. The crucible was heated to 1573 K and dwelt for 10 h and then slowly cooled down to room temperature. After that the single crystals were separated from the flux by washing with deionized water. The obtained single crystal had typical dimensions 0.8 × 0.8 × 0.3 mm$^3$. The experimental XRD pattern of Sr$_3$Ir$_2$O$_7$ and calculated pattern based on the standard ICSD (075587) date are shown in the Supplementary Material in Fig. S4a. This result indicated that our sample is tetragonal phase (space group I4/mmm) at ambient condition and has good quality.

### High-pressure Raman measurements
Raman spectra were collected on a single-crystal Sr$_3$Ir$_2$O$_7$ at beamline 22-IR-1 of the National Synchrotron Light Source II, Brookhaven National laboratory. The sample was loaded inside a symmetric-type diamond anvil cell (DAC) and the DAC was placed in a microscopy cryostat system (Cryo Industries of America, Inc.). A pair of ultralow fluorescence type II diamonds with culet size ~300 μm were used. A Spectra-Physics Excelsior solid-state laser with a wavelength of 532 nm was used in the Raman measurement. Potassium bromide (KBr) was used as the pressure transmitting medium, and the pressure inside the cell was determined by the shift of the ruby fluorescence line. The laser power was less than 1 mW. In our measurements, 300 grooves/mm grating and ~1–3 μm beam spot were applied.

## High-pressure transport measurements

Electrical resistance was measured with a standard four-probe-electrode circuit on a single-crystal $Sr_3Ir_2O_7$. A T301 stainless steel gasket with cubic boron nitride/epoxy mixture powder inserts was used. Si oil was used as a pressure medium and the pressure was determined using ruby fluorescence technique. Four thin gold probes were attached to the samples with silver glue to measure the resistance.

## High-pressure synchrotron diffraction measurements

The in situ high-pressure XRD measurements were carried out on a single-crystal $Sr_3Ir_2O_7$ at beamline 16-BM-D of the Advanced Photon Source (APS), Argonne National Laboratory. A symmetric type DAC with culet size of 300 μm was used for the measurement. Neon was used as a pressure medium and the pressure was determined using ruby fluorescence technique. The incident monochromatic x-ray beam energy was set to 30 keV ($\lambda = 0.4133$ Å). Diffraction patterns were recorded on a MAR345 image plate. We aligned the sample to the vertical-rotation axis and collected diffraction patterns in a step-scan method (1.5 s/step) with 1.0° step over the range from $-20°$ to $20°$ up to 33.2 GPa, similar to the "rotation method" used in conventional single crystal crystallography.

## Model Hamiltonian and spin-wave theory

We describe the pressure evolution of magnon excitation using the spin model reported in the ref.[22] Below we review the interaction terms in the model, which includes both intralayer ($H_{ab}$) and interlayer ($H_c$) Hamiltonians:

$$H_{ab} = \sum_{<i,j>}\left[JS_i \cdot S_j + \Gamma S_i^z S_j^z + D(S_i^x S_j^y - S_i^y S_j^x)\right] + \sum_{<<i,j>>} J_2 S_i \cdot S_j + \sum_{<<<i,j>>>} J_3 S_i \cdot S_j$$

$$H_c = \sum_{i}\left[J_c S_i \cdot S_{i+z} + \Gamma_c S_i^z S_{i+z}^z + D_c(S_i^x S_{i+z}^y - S_i^y S_{i+z}^x)\right] + \sum_{<i,j>} J_{2c} S_i \cdot S_{j+z}$$

where $<i,j>$, $<<i,j>>$, and $<<<i,j>>>$ denote the first, second, and third nearest neighbors within the a-b plane. $J$, $J_2$, and $J_3$ represent the isotropic coupling constants. The anisotropic coupling term $\Gamma$ stems from Hund's exchange interaction and staggered rotations of octahedra. The latter also results in a Dzyaloshinsky-Moriya (DM) interaction,[22] characterized by the constant $D$. For the nearest-interlayer interactions, $J_c$, $\Gamma_c$, and $D_c$ were adopted for the similar coupling constants along the c-axis, while $J_{2c}$ stands for the next-nearest-neighbor interlayer coupling. All these isotropic and anisotropic exchange coupling constants (except for the long-range interactions $J_2$, $J_3$, and $J_{2c}$) can be expressed in terms of the three microscopic parameters: the IrO$_6$ rotation angle $\alpha$, the effective tetragonal distortion $\theta$, and the ratio of Hund's coupling to onsite Coulomb interaction $\eta = (J_H/U)$. Here, we use the parameters in the reference for ambient pressure. We also assume $J_{2c} = 0.25\ J_c$ and keep $J_2 = 11.9$ meV, $J_3 = 14.6$ meV, and $\eta$ as constants.[22]

Within linear spin-wave theory (applicable in the antiferromagnetic phase), the single-magnon excitation energy in the Brillouin zone center (i.e., that observed by Raman measurements) is expressed by $\omega = 1/2\sqrt{(4\Gamma + \Gamma_c)(8J + 2J_c + 4\Gamma + \Gamma_c) - (4D + D_c)^2}$. Our calculations reveal that the major influence of pressure is to reduce the anisotropic coupling $\Gamma$, e.g., from 4.4 meV at ambient pressure to 2.0 meV at around 20 GPa. Also, both $J_c$ and $D_c$ are enhanced, while other coupling constants are only slightly affected. We notice that tilting of the IrO$_6$ octahedra along the c-axis can reduce the interlayer coupling $J_c$ and thus soften the magnon energy, although overall $J_c$ is enhanced due to a pressure-enhanced bandwidth.

## DATA AVAILABILITY

The data that support the findings of this study are available from the corresponding author upon reasonable request.

## ACKNOWLEDGEMENTS

This work is supported by National Key R&D Program of China No. 2018YFA0305703. The x-ray diffraction measurements were performed at sectors 16 BM-D of the Advanced Photon Source, a U.S. Department of Energy (DOE) Office of Science user facility operated by Argonne National Laboratory (ANL) supported by the U.S. DOE Award No. DE-AC02-06CH11357. The Raman experiments were performed at beamline 22-IR-1 of the National Synchrotron Light Source II (NSLS-II), Brookhaven National Laboratory, supported by NSF (Cooperative Agreement EAR 1606856, COMPRES) and DOE/NNSA (DE-NA-0002006, CDAC). NSLS-II is supported by the DOE/BES (DE-SC0012704). The electric transport measurements were performed at the Center for High-Pressure Science and Technology Advanced Research. The authors thank S. Tkachev for help with the gas loading at the Advanced Photon Source. Y.D. and H.-k. M. acknowledges the support from DOE-BES under Award No. DE-FG02-99ER45775 and NSFC Grant No. U1530402. This work is also supported by National Key R&D Program of China No. 2016YFA0300604, 2017YFA0302901, and the National Natural Science Foundation of China No. 11774399, 11474330, 91750111, and Science Challenge Project, No. TZ2016001, and the Fundamental Research Funds for the Central Universities, China, No. GK201801009.

## AUTHOR CONTRIBUTIONS

Y.D. and J.Z. designed the project. J.Z., H.D., and Z.L. performed the Raman measurements. J.Z., S.Y., and H.D. carried out the XRD and transport measurements. C.Y., D.Y., and Y.S. synthesized the single crystals. J.C. performed the theoretical calculations. J.Z., Y.D., and J.C. analyzed the data. Y.D. supervised the project. All the authors helped with the project and read and commented on the manuscript.

## ADDITIONAL INFORMATION

**Supplementary information** accompanies the paper on the npj Quantum Materials website (https://doi.org/10.1038/s41535-019-0162-3).

**Competing interests:** The authors declare no competing interests.

**Publisher's note:** Springer Nature remains neutral with regard to jurisdictional claims in published maps and institutional affiliations.

## REFERENCES

1. Wang, F. & Senthil, T. Twisted Hubbard model for $Sr_2IrO_4$: magnetism and possible high temperature superconductivity. Phys. Rev. Lett. **106**, 136402 (2011).
2. Kitagawa, K. et al. A spin-orbital-entangled quantum liquid on a honeycomb lattice. Nature **554**, 341–345 (2018).
3. Wan, X., Turner, A. M., Vishwanath, A. & Savrasov, S. Y. Topological semimetal and Fermi-arc surface states in the electronic structure of pyrochlore iridates. Phys. Rev. B **83**, 205101 (2011).
4. Price, C. C. & Perkins, N. B. Critical properties of the Kitaev-Heisenberg model. Phys. Rev. Lett. **109**, 187201 (2012).
5. Chaloupka, J., Jackeli, G. & Khaliullin, G. Zigzag magnetic order in the iridium oxide $Na_2IrO_3$. Phys. Rev. Lett. **110**, 097204 (2013).
6. Watanabe, H., Shirakawa, T. & Yunoki, S. Monte Carlo study of an unconventional superconducting phase in iridium oxide $J_{eff}$=1/2 Mott insulators induced by carrier doping. Phys. Rev. Lett. **110**, 027002 (2013).
7. Yonezawa, S., Muraoka, Y., Matsushita, Y. & Hiroi, Z. Superconductivity in a pyrochlore-related oxide $KOs_2O_6$. J. Phys. Condens. Matter **16**, L9–L12 (2004).
8. Jackeli, G. & Khaliullin, G. Mott insulators in the strong spin-orbit coupling limit: from Heisenberg to a quantum compass and Kitaev models. Phys. Rev. Lett. **102**, 017205 (2009).
9. Witczak-Krempa, W., Chen, G., Kim, Y. B. & Balents, L. Correlated quantum phenomena in the strong spin-orbit regime. Annu. Rev. Conden. Ma. P. **5**, 57–82 (2014).
10. Cao, G. & Schlottmann, P. The challenge of spin-orbit-tuned ground states in iridates: a key issues review. Rep. Prog. Phys. **81**, 042502 (2018).
11. Cao, G. et al. Electrical control of structural and physical properties via strong spin-orbit interactions in $Sr_2IrO_4$. Phys. Rev. Lett. **120**, 017201 (2018).
12. Liu, H. & Khaliullin, G. Pseudo Jahn-Teller effect and magnetoelastic coupling in spin-orbit Mott insulators. Phys. Rev. Lett. **122**, 057203 (2019).
13. Plotnikova, E. M., Daghofer, M., van den Brink, J. & Wohlfeld, K. Jahn-Teller effect in systems with strong on-site spin-orbit coupling. Phys. Rev. Lett. **116**, 106401 (2016).
14. Samanta, K., Ardito, F. M., Souza-Neto, N. M. & Granado, E. First-order structural transition and pressure-induced lattice/phonon anomalies in $Sr_2IrO_4$. Phys. Rev. B **98**, 094101 (2018).
15. Ding, Y. et al. Novel high-pressure monoclinic metallic phase of $V_2O_3$. Phys. Rev. Lett. **112**, 056401 (2014).
16. Hogan, T. et al. First-order melting of a weak spin-orbit Mott insulator into a correlated metal. Phys. Rev. Lett. **114**, 257203 (2015).
17. Cao, G. et al. Anomalous magnetic and transport behavior in the magnetic insulator $Sr_3Ir_2O_7$. Phys. Rev. B **66**, 214412 (2002).
18. Subramanian, M. A., Crawford, M. K. & Harlow, R. L. Single crystal structure determination of double layered strontium iridium oxide $Sr_3Ir_2O_7$. Mater. Res. Bull. **29**, 645–650 (1994).

19. Matsuhata, H. et al. Crystal structure of $Sr_3Ir_2O_7$ investigated by transmission electron microscopy. *J. Solid. State Chem.* **177**, 3776–3783 (2004).
20. Hogan, T. et al. Structural investigation of the bilayer iridate $Sr_3Ir_2O_7$. *Phys. Rev. B* **93**, 134110 (2016).
21. Gretarsson, H. et al. Two-magnon raman scattering and pseudospin-lattice interactions in $Sr_2IrO_4$ and $Sr_3Ir_2O_7$. *Phys. Rev. Lett.* **116**, 136401 (2016).
22. Kim, J. et al. Large spin-wave energy gap in the bilayer iridate $Sr_3Ir_2O_7$: evidence for enhanced dipolar interactions near the Mott metal-insulator transition. *Phys. Rev. Lett.* **109**, 157402 (2012).
23. Moretti Sala, M. et al. Evidence of quantum dimer excitations in $Sr_3Ir_2O_7$. *Phys. Rev. B* **92**, 024405 (2015).
24. Cetin, M. F. Light scattering in spin orbit coupling dominated systems. Ph.D thesis, Technische Universitaet Carolo-Wilhelmina zu Braunschweig (2012).
25. Zhao, Z. et al. Pressure induced second-order structural transition in $Sr_3Ir_2O_7$. *J. Phys. Condens. Matter* **26**, 215402 (2014).
26. Li, L. et al. Tuning the $J_{eff}=1/2$ insulating state via electron doping and pressure in the double-layered iridate $Sr_3Ir_2O_7$. *Phys. Rev. B* **87**, 235127 (2013).
27. Chu, H. et al. A charge density wave-like instability in a doped spin-orbit-assisted weak Mott insulator. *Nat. Mater.* **16**, 200–203 (2017).
28. Ding, Y. et al. Pressure-induced confined metal from the Mott insulator $Sr_3Ir_2O_7$. *Phys. Rev. Lett.* **116**, 216402 (2016).
29. Donnerer, C. et al. Pressure dependence of the structure and electronic properties of Sr3Ir2O7. *Phys. Rev. B* **93**, 174118 (2016).
30. Kim, J. W. et al. Dimensionality driven spin-flop transition in layered iridates. *Phys. Rev. Lett.* **109**, 037204 (2012).
31. Qi, T. F. et al. Spin-orbit tuned metal-insulator transitions in single-crystal $Sr_2Ir_{1-x}Rh_xO_4 (0\leq x\leq 1)$. *Phys. Rev. B* **86**, 125105 (2012).
32. Cetin, M. F. et al. Crossover from coherent to incoherent scattering in spin-orbit dominated $Sr_2IrO_4$. *Phys. Rev. B* **85**, 195148 (2012).
33. Zocco, D. A. et al. Persistent non-metallic behavior in $Sr_2IrO_4$ and $Sr_3Ir_2O_7$ at high pressures. *J. Phys. Condens. Matter* **26**, 255603 (2014).
34. Yamaura, J. et al. Tetrahedral magnetic order and the metal-insulator transition in the pyrochlore lattice of $Cd_2Os_2O_7$. *Phys. Rev. Lett.* **108**, 247205 (2012).
35. Sohn, C. H. et al. Strong spin-phonon coupling mediated by single ion anisotropy in the all-in-all-out pyrochlore magnet $Cd_2Os_2O_7$. *Phys. Rev. Lett.* **118**, 117201 (2017).
36. Tardif, S. et al. All-in-all-out magnetic domains: x-ray diffraction imaging and magnetic field control. *Phys. Rev. Lett.* **114**, 147205 (2015).
37. Shinaoka, H., Miyake, T. & Ishibashi, S. Noncollinear magnetism and spin-orbit coupling in 5d pyrochlore oxide $Cd_2Os_2O_7$. *Phys. Rev. Lett.* **108**, 247204 (2012).
38. Bogdanov, N. A. et al. Magnetic state of pyrochlore $Cd_2Os_2O_7$ emerging from strong competition of ligand distortions and longer-range crystalline anisotropy. *Phys. Rev. Lett.* **110**, 127206 (2013).
39. Ge, M. et al. Lattice-driven magnetoresistivity and metal-insulator transition in single-layered iridates. *Phys. Rev. B* **84**, 100402 (2011).
40. Haskel, D. et al. Pressure tuning of the spin-orbit coupled ground state in $Sr_2IrO_4$. *Phys. Rev. Lett.* **109**, 027204 (2012).